\documentclass{article}

\usepackage[utf8]{inputenc}     
\usepackage[margin=10pt,font=small,labelfont=bf,labelsep=endash]{caption}
\usepackage[T1]{fontenc}
\usepackage{setspace}
\usepackage{booktabs}
\usepackage{multirow}
\usepackage{tabularx}
\usepackage{paralist}
\usepackage{ifthen}
\usepackage{listings}
\usepackage{amsmath}
\usepackage{amsfonts}
\usepackage{amssymb}
\usepackage{array}
\usepackage{color}
\usepackage[table]{xcolor}
\usepackage{algorithm}
\usepackage{algpseudocode}
\usepackage{url}
\usepackage{rotating}
\usepackage{subfig}
\usepackage{arydshln}
\usepackage{todonotes}
\usepackage[normalem]{ulem}
\usepackage{siunitx}
\usepackage{diagbox}
\usepackage[htt]{hyphenat}
\usepackage{enumitem}
\usepackage{tikz}
\usepackage{pgfplots}
\usetikzlibrary{plotmarks}
\usepackage{authblk} 
\usepackage{comment} 
\usepackage{xspace} 
\usepackage[top=3cm, bottom=3cm, inner=2.5cm, outer=2.5cm]{geometry}
\usepackage[absolute]{textpos}

\RequirePackage[sectionbib,round]{natbib}
\RequirePackage{fancyvrb}
\RequirePackage{alltt}
\DefineVerbatimEnvironment{example}{Verbatim}{}
\renewenvironment{example*}{\begin{alltt}}{\end{alltt}}

\providecommand{\href}{\url}
\providecommand{\address}{}
\providecommand{\email}{}

\graphicspath{{../thesis/img_out/}{../thesis/img_other/}{images/}}
\makeatletter
\def\input@path{{../thesis/}}
\makeatother

\excludecomment{rjonly}
\includecomment{arxivonly}
\newcommand{\forarxivonly}[1]{}
\begin{arxivonly}
\renewcommand{\forarxivonly}[1]{#1}
\end{arxivonly}

\providecommand{\nameref}{\ref}

\newcommand{\progfunc}{\textnhtt}
\newcommand{\swpackage}{\texttt}

\providecommand{\pkg}{\swpackage}
\providecommand{\code}{\progfunc}
\providecommand{\CRANpkg}{\progfunc}

\floatstyle{ruled}
\newfloat{proc}{thp}{lop}
\floatname{proc}{Procedure}

\makeatletter
\@ifpackageloaded{algorithm}
	{\algrenewcommand\algorithmicforall{\textbf{for each}}}
	{}
\@ifpackageloaded{amsmath}{
	
	}
	{}
\makeatother

\newcommand*\mean[1]{\overline{#1}}

\newcommand{\TikZ}{Ti\textit{k}Z\xspace}

\begin{document}

\date{}

\begin{textblock*}{202mm}(3mm,3mm)
\noindent The peer-reviewed version of this paper is published in The R Journal (\url{https://journal.r-project.org/archive/2016-2/fachada-rodrigues-lopes-etal.pdf}). This version is typeset by the authors and differs only in pagination and typographical detail.
\end{textblock*}


\begin{rjonly}

\title{\pkg{micompr}: An R Package for Multivariate Independent Comparison of Observations}

\author{by Nuno Fachada, Jo\~{a}o Rodrigues, Vitor V. Lopes, Rui C. Martins and Agostinho C. Rosa}

\end{rjonly}

\begin{arxivonly}

\title{\pkg{micompr}: An R Package for Multivariate Independent Comparison of Observations}

\author[1]{Nuno Fachada}
\author[2]{Jo\~{a}o Rodrigues}
\author[3]{Vitor V. Lopes}
\author[4]{Rui C. Martins}
\author[1]{Agostinho C. Rosa}

\affil[1]{Institute for Systems and Robotics, LARSyS, Instituto Superior T\'{e}cnico, Universidade de Lisboa, Lisboa, Portugal}
\affil[2]{\'{E}cole Polytechnique F\'{e}d\'{e}rale de Lausanne, Lausanne, Switzerland}
\affil[3]{UTEC - Universidad de Ingenier\'{i}a \& Tecnolog\'{i}a, Lima, Per\'{u}}
\affil[4]{Life and Health Sciences Research Institute, School of Health Sciences, University of Minho, Braga, Portugal}

\end{arxivonly}

\maketitle

\abstract{The R package \pkg{micompr} implements a procedure for assessing if two or more multivariate samples are drawn from the same distribution. The procedure uses principal component analysis to convert multivariate observations into a set of linearly uncorrelated statistical measures, which are then compared using a number of statistical methods. This technique is independent of the distributional properties of samples and automatically selects features that best explain their differences. The procedure is appropriate for comparing samples of time series, images, spectrometric measures or similar high-dimension multivariate observations. }

\section{Introduction}\label{micompr:sec:intro}

The aim of this paper is to present the \CRANpkg{micompr} package for R\forarxivonly{ \citep{r2015stats}}, which implements a procedure for comparing multivariate samples associated with different factor levels or groups. The research goal is to differentiate among pre-specified, well-defined classes or groups of sampling entities generating highly multivariate observations in which the dimensions or dependent variables are correlated, and to test for significant differences among groups. The procedure uses principal component analysis (PCA) \citep{jolliffe2002principal} to convert multivariate observations into a set of linearly uncorrelated statistical measures, which are then compared using a number of statistical methods, such as hypothesis tests and score plots.

This technique has several desirable attributes: a) it automatically selects observation features that best explain sample differences; b) it does not depend on the distributional properties of samples; and, c) it simplifies the researchers' work, as it can be used directly on multivariate observations. The procedure is appropriate for comparing samples of multivariate observations with highly correlated and similar scale dimensions, such as time series, images or spectrometric measures. However, the \pkg{micompr} package goes one step further by also accommodating the simultaneous comparison of multiple observation types, i.e., multiple \emph{outputs} from a given ``system''. In this context, a ``system'' can be defined as an abstract entity capable of generating one or more stochastic data streams, i.e., \emph{outputs}. Thus, \pkg{micompr} can determine if two or more instances of such a ``system'' display the same behavior by comparing observations of their \emph{outputs}. 

The remainder of this paper is organized as follows. First, in Section \nameref{micompr:sec:finding}, commonly used techniques for differentiating samples of multivariate observations are discussed. The methodology employed by \pkg{micompr} is described in Section \nameref{micompr:sec:theory}. Section \nameref{micompr:sec:micompr} introduces the software and its architecture, namely the available objects and functions. Several concrete application examples, and how the ``system''-\emph{output} terminology fits each one, are presented in Section \nameref{micompr:sec:examples}. The paper closes with Section \nameref{micompr:sec:summary}, in which the overall approach and the R package are summarized.

\section{Testing for significant differences in multivariate samples}\label{micompr:sec:finding}

Two-sample or multi-sample hypothesis tests are commonly used for assessing statistically dissimilarity in univariate samples, i.e., samples composed of scalar observations. If samples are drawn from normally distributed populations, the $t$ (two samples) and ANOVA ($n$-samples) tests are adequate \citep{montgomery2010applied}. Non-parametric tests are more appropriate if population normality cannot be assumed. The Mann-Whitney $U$ test \citep{gibbons2011nonparametric} and the Kolmogorov-Smirnov test \citep{massey1951kolmogorov} are typically employed for comparing two samples. The Kruskal-Wallis test \citep{kruskal1952use} extends the former for the $n$-sample case.

Multivariate analysis of variance (MANOVA) \citep{krzanowski1998,tabachnick2013using} can be used as a statistical test for comparing multivariate samples. In this context, samples are composed of multi-dimensional observations, for which each dimension is a dependent variable (DV). However, MANOVA is not appropriate for cases with highly correlated DVs and when the number of DVs or dimensions is higher than the number of observations. Additionally, MANOVA is a parametric method which makes a series of assumptions on the underlying data which are not always met in practice.

Analogous non-parametric tests exist, but they are not as widespread and are commonly oriented towards specific research topics. Multiple Response Permutation Procedures (MRPP) \citep{mielke1976multi} and associated permutation-based methods, such as ANOSIM \citep{clarke1993nonparametric} or permutational MANOVA \citep{anderson2001new}, test for differences in distances between observations from each group. These tests are implemented in the \CRANpkg{vegan} package \citep{vegan2016}, typically used in Ecology studies. The \CRANpkg{Blossom} package \citep{talbert2016blossom} also provides MRPP and other distance-function based permutation tests. In a similar note, \cite{szekely2004testing} proposed a multi-sample test for equality of multivariate distributions based on the Euclidean distance between sample elements. The test statistic belongs to a class of multivariate statistics (energy statistics) proposed by the same authors. The \CRANpkg{energy} package \citep{rizzo2016energy} implements this test and other energy statistics-related functionality. The cross-match test is another distance-based test \citep{rosenbaum2005exact}, with the particularity of not requiring permutation techniques. It is available for the R environment via the \CRANpkg{crossmatch} package \citep{heller2012crossmatch}. In turn, the \CRANpkg{cramer} package \citep{franz2014cramer} provides a multivariate implementation of the non-parametric two-sample Cram\'{e}r test, originally proposed by \cite{baringhaus2004new}. The critical value of the test can be determined with bootstrap (ordinary or permutation-based) or eigenvalue-based methods. Another test which avoids permutation was proposed by \cite{duong2012closed}. It is a kernel-based test, originally developed to assess the statistical differences between two cellular topologies. The test is implemented in the \CRANpkg{ks} package \citep{duong2016ks}, although limited to six-dimensional data.

An alternative to multivariate tests is to extract a number of statistical summaries (e.g., averages or extreme values) or specific points from individual multivariate observations, and then perform a univariate test for each summary measure. This approach also has its issues: a) it does not provide a single answer, i.e., it will yield as many $p$-values as there are summary measures; b) the choice of summary is problem-dependent and empirically driven, and consequently, error-prone, in the sense that the chosen summaries may not be representative of the original multivariate observations. While only careful analysis can minimize the latter issue, the former problem can be addressed with a multiple comparison adjustment procedure, such as the Bonferroni correction \citep{shaffer1995multiple}.

\section{Independent comparison of multivariate observations}\label{micompr:sec:theory}

Given a matrix $\mathbf{X}_{(n \times m)}$ of $n$ observations and $m$ variables or dimensions, PCA can be used to obtain matrix $\mathbf{T}_{(n \times r)}$, which is the representation of $\mathbf{X}_{(n \times m)}$ in the principal components (PCs) space, and vector $\boldsymbol{\lambda}_{(1 \times r)}$, containing the eigenvalues of the covariance matrix of the original mean-centered data. Rows of $\mathbf{T}$ directly correspond to the observations of the original samples, while columns correspond to PCs. Columns are ordered by decreasing variance, i.e., the first column corresponds to the first PC, and so on. Variance is given by the eigenvalues in vector $\boldsymbol{\lambda}$, which are likewise ordered, each eigenvalue corresponding to the variance of the columns of $\mathbf{T}$. The percentage of variance explained by each PC can be obtained by dividing the respective eigenvalue with the sum of all eigenvalues. At this stage, PCA-reshaped observations associated with different groups can be compared using statistical methods. More specifically, hypothesis tests can be used to check if the sample projections on the PC space are drawn from populations with the same distribution. There are two possible lines of action:

\begin{enumerate}

\item Apply a MANOVA test to the samples, where each observation has $q$-dimensions, corresponding to the first $q$ PCs (dimensions) such that these explain a user\hyp{}defined minimum percentage of variance.

\item Apply a univariate test to observations in individual PCs. Possible tests include the $t$-test and the Mann-Whitney U test for comparing two samples, or ANOVA and Kruskal-Wallis test, which are the respective parametric and non-parametric versions for comparing more than two samples.
\end{enumerate}

The MANOVA test yields a single $p$-value from the simultaneous comparison of observations along multiple PCs. An equally succinct answer can be obtained with a univariate test using the Bonferroni correction or a similar method for handling $p$-values from multiple comparisons. However, both approaches will not prioritize dimensions, even though the first PCs can be more important for characterizing an output, as they explain more variance. In the univariate case one can prioritize PCs according to the explained variance using the weighted Bonferroni procedure \citep{rosenthal1983ensemble}.  

Conclusions concerning whether samples are statistically similar can be drawn by analyzing the $p$-values produced by the employed statistical tests, which should be below the typical 1\% or 5\% when samples are significantly different. In such case, less PCs should be required to explain the same percentage of variance than when, in the same context, no significant differences are found. The scatter plot of the first two PC dimensions can also provide visual, although subjective feedback on sample similarity.

While the procedure is most appropriate for comparing multivariate observations with highly correlated and similar scale dimensions, assessing the similarity of ``systems'' with multiple outputs of different scales is also possible. The simplest approach would be to apply the proposed method to samples of individual outputs, and analyze the results in a multiple comparison context. An alternative approach consists in concatenating, observation-wise, all outputs, centered and scaled to the same order of magnitude, thus reducing a ``system'' with $k$ outputs to a ``system'' with one output. The proposed method would then be applied to samples composed of concatenated observations encompassing the existing outputs. This technique is described in detail by \citet{fachada2015model} in the context of comparing simulation outputs.

\section{The \pkg{micompr} package}\label{micompr:sec:micompr}

\subsection{Overview}

The \pkg{micompr} package for the R statistical computing environment implements the methodology proposed in Section \nameref{micompr:sec:theory}. Here we describe version 1.0.1 of the package, which is available at \url{https://cran.r-project.org/package=micompr}. The development version is hosted at \url{https://github.com/fakenmc/micompr}. The package is covered by the MIT license.

The \pkg{micompr} package is built upon two functions, \code{cmpoutput} and \code{micomp}. The former compares two or more samples of multivariate observations collected from one output. The latter is used for comparing multiple outputs and/or comparing outputs in multiple contexts. \code{grpoutputs} is a helper function for loading data from two or more set of files and preparing the data to be processed by the \code{cmpoutput} and/or \code{micomp} functions. \code{assumptions} is a generic function for assessing the assumptions of the parametric tests used in sample comparisons.

\subsection{Architecture}

\pkg{micompr} is structured according to the S3 object-oriented system. The \code{cmpoutput}, \code{micomp} and \code{grpoutputs} functions produce S3 objects with the same name. The package also provides the generic function \code{assumptions}, and two concrete implementations of methods for ''cmpoutput'' and ''micomp'' objects, which return objects of class ''assumptions\_cmpoutput'' and ''assumptions\_micomp'', respectively. All classes have method implementations of the common S3 generic functions \code{print}, \code{summary} and \code{plot}. Additionally, method implementations of the \code{toLatex} function, for producing user-configurable \LaTeX\ tables with information about the performed comparisons, are provided for ''cmpoutput'' and ''micomp'' objects.

\subsubsection{\code{grpoutputs}}

This function groups outputs from sets of files containing multiple observations into samples. It returns a list of output matrices, ready to be processed by \code{micomp}. Alternatively, individual output matrices can be handled by \code{cmpoutput}. Separate files contain one multivariate observation of one or more outputs, one column per output, one row per dimension or variable. Each specified set of files is associated with a different sample. The function is also able to create an additional concatenated output, composed from the centered and scaled original outputs.

The plot method for ``grpoutputs'' objects shows $k$ plots, one per output. Output observations are plotted on top of each other, with different samples colored distinctively. The \code{summary} method for ``grpoutputs'' objects returns a list containing two elements: a) the $n \times m$ dimensions of each output matrix; and, b) the sizes of individual samples. The \code{print} method for ``grpoutputs'' objects simply outputs the summary in a more adequate presentation format.

\subsubsection{\code{cmpoutput}}

The \code{cmpoutput} function is at the core of \pkg{micompr}. It compares two or more samples of multivariate observations using the technique described in Section \nameref{micompr:sec:theory}. It accepts an output matrix, $\mathbf{X}_{(n \times m)}$, with $n$ observations and $m$ variables or dimensions, a factor vector of length $n$, specifying the group associated with each observation, and a vector of explained variances with which to determine the number of PCs to use in the MANOVA test (alternatively, the number of PCs can also be directly specified). The function returns matrix $\mathbf{T}_{(n \times r)}$ of PCA scores and the $p$-values for the performed statistical tests, namely: a) a MANOVA test for each explained variance (or number of PCs); and, b) parametric ($t$-test or ANOVA) and non-parametric (Mann-Whitney or Kruskal-Wallis) univariate tests for each PC. Regarding the latter, the function also returns $p$-values adjusted with the weighted Bonferroni correction, using the percentages of explained variance by PC as weights.

The \code{plot} implementation for ``cmpoutput'' objects shows six sub-plots, namely a scatter plot with the PC1 vs. PC2 scores and five bar plots. The horizontal scale of the latter consists of the $r$ PCs, and the vertical bars represent the explained variance (one plot) or univariate parametric and non-parametric $p$-values, before and after weighted Bonferroni correction (four plots). The \code{summary} method for ``cmpoutput'' objects returns a list with the following items: a) percentage of variance explained by each PC; b) $p$-values of the MANOVA test or tests; c) $p$-values of the parametric test, per PC, before and after weighted Bonferroni correction; d) $p$-values of the non-parametric test, per PC, before and after weighted Bonferroni correction; and, e) name of the parametric and non-parametric univariate tests employed (either $t$-test and Mann-Whitney $U$ test for comparing two samples, or ANOVA and Kruskal-Wallis for more than two samples). The \code{print} method for ``cmpoutput'' objects shows the information provided by the \code{summary} implementation, but the $p$-values of the univariate tests are only shown for the first PC.

\subsubsection{\code{micomp}}

The \code{micomp} function performs one or more comparisons of multiple outputs, invoking \code{cmpoutput} for each comparison/output combination. It accepts a list of comparisons, where individual comparisons can have one of two configurations: a) a vector of folders and a vector of file sets containing data in the format required by \code{grpoutputs}, where each file set corresponds to a different sample; and, b) a ``grpoutputs'' object, passed directly. The returned objects, of class ``micomp'', are basically two-dimensional lists of ``cmpoutput'' instances, with rows associated with individual outputs, and columns with separate comparisons.

The \code{plot} method for ``micomp'' objects shows the PC1 vs. PC2 score plots for each comparison/output combination. The \code{summary} implementation for ``micomp'' objects returns a list of comparisons, each one containing a $a \times k$ matrix of $p$-values or number of PCs, associated with $a \geq 6$ measures and $k$ outputs. Four rows represent the $p$-values of the parametric and non-parametric univariate tests for the first PC, before and after weighted Bonferroni correction. The remaining pairs of rows are associated with the MANOVA test for a given percentage of variance to explain. One row shows the $p$-values, and the other displays the number of PCs required to explain the specified percentage of variance for the given output. As with other \pkg{micompr} objects, the \code{print} method for ``micomp'' objects also shows the summary with a better presentation.

\subsubsection{\code{assumptions}}

\code{assumptions} is a generic function which performs a number of statistical tests concerning the assumptions of the parametric tests performed by the package functions. Implementations of this generic function exist for ``cmpoutput'' and ``micomp'' objects. The former method returns objects of class ``assumptions\_cmpoutput'' containing results of the assumptions tests for a single output comparison. The latter returns a two-dimensional list of ``assumptions\_cmpoutput'' objects, with rows associated to individual outputs, and columns to separate comparisons. These objects are tagged with the ``assumptions\_micomp'' class attribute.

The following assumptions are checked: a) observations are normally distributed within each sample along individual PCs (Shapiro-Wilk test); b) observations follow a multivariate normal distribution within each sample for all PCs used in MANOVA (Royston test); c) samples have homogeneous variance along individual PCs (Bartlett test); and, d) samples have homogeneous covariance matrices for all PCs used in MANOVA (Box's $M$ test). Assumptions a) and c) should be verified for the parametric test applied to each PC, while assumptions b) and d) should be verified for individual MANOVA tests performed for each variance to explain (or, alternatively, for each specified number of PCs).

The \code{plot} implementations for classes ``assumptions\allowbreak{}\_\allowbreak{}cmpoutput'' and ``assumptions\allowbreak{}\_\allowbreak{}micomp'' display a number of bar plots for the $p$-values of the performed tests. These are more detailed for ``assumptions\allowbreak{}\_\allowbreak{}cmpoutput'' objects, showing the $p$-values of the univariate test for all PCs. For ``assumptions\_micomp'' objects, one bar plot is shown per output/comparison combination, but in the case of the univariate tests only the $p$-values of the first PC are shown. Implementations of \code{summary} return a list of tabular data containing the $p$-values of the assumption tests. The \code{summary} method for ``assumptions\allowbreak{}\_\allowbreak{}cmpoutput'' objects returns a list with two matrices of $p$-values, one for the MANOVA tests, another for the univariate tests. The \code{summary} method for ``assumptions\allowbreak{}\_\allowbreak{}micomp'' objects follows the approach taken by the \code{summary} method for ``micomp'' objects, returning a list of $p$-value matrices, one matrix per comparison. Rows of individual matrices correspond to the assumptions tests, and columns to outputs. The \code{print} methods for ``assumptions\allowbreak{}\_\allowbreak{}cmpoutput'' objects and for ``assumptions\allowbreak{}\_\allowbreak{}micomp'' objects again show the summary information in a printable format.

\subsubsection{\code{toLatex} methods for ``cmpoutput'' and ``micomp'' objects}

These methods are implementations of the \code{toLatex} generic function, and convert ``cmpoutput'' and ``micomp'' objects to character vectors representing \LaTeX\ tables. The generated tables are configurable via function arguments, with sensible defaults. Tables can present the following data for each output/comparison combination: a) number of principal components required to explain a user-specified percentage of variance; b) MANOVA $p$-value for a user-specified percentage of variance to explain or number of PCs; c) parametric test $p$-value for a given PC, before and/or after weighted Bonferroni correction; d) non-parametric test $p$-value for a given PC, before and/or after weighted Bonferroni correction; e) variance explained by a specific PC; and, f) a score plot with the output projection on the first two PCs.

\subsubsection{Other functions}

The \pkg{micompr} package is bundled with additional functions whose purpose is to aid the main package methods do their job. However, some of these may be useful in other contexts.

The \code{concat\_outputs} function concatenates outputs collected from multiple observations. It accepts two arguments, namely a list of output matrices, and the centering and scaling method. Several centering and scaling methods, such as ``range'', ``iqrange'', ``vast'' or ``pareto'' \citep{berg2006centering}, are recognized in the second argument. The function returns an $n \times p$ matrix of $n$ observations with length $p$, which is the sum of individual output lengths. Lower-level centering and scaling of individual outputs is performed by the \code{centerscale} function, which accepts a numeric vector and returns a new vector, centered and scaled with the specified method.

The \code{pvalf} generic function formats $p$-values for \LaTeX . A concrete default implementation is used by the \pkg{micompr} \code{toLatex} implementations. This implementation underlines and double-underlines $p$-values lower than $0.05$ and $0.01$, respectively, although these limits are configurable, and underlining can be turned off by setting both limits to zero. It is also possible to specify a limit below which $p$-values are capped. For example, if this limit is set to $1 \times 10^{-5}$, a $p$-value equal to $1 \times 10^{-6}$ would be displayed as ``$< 1\text{e}^{-5}$''. The default method of \code{pvalf} will format $p$-values lower than $5 \times 10^{-4}$ using scientific E notation, which is more compact and thus a better fit for tables. $p$-values between $5 \times 10^{-4}$ and $1$ are formatted using regular decimal notation with three decimal places. This aspect is not configurable. However, another implementation of \code{pvalf} can be passed to the \pkg{micompr} \code{toLatex} implementations if different formatting is desired. 

Simple \TikZ 2D scatter plots, as the ones produced by the \pkg{micompr} \code{toLatex} implementations, can be generated with the \code{tikzscat} function. The function accepts the data to plot, an $n$ x 2 numeric matrix, of $n$ observations and $2$ dimensions, and a factor vector specifying the levels or groups associated with each observation. Several plot characteristics, such as mark types, scale and axes color, are configurable via function arguments. \code{tikzscat} returns a string containing the \TikZ figure code for plotting the specified data.

\subsection{Included data}

The package includes test data produced by several implementations of the Predator-Prey for High Performance Computing (PPHPC) simulation model \citep{fachada2015template}. The data is provided in \code{rdata} format, and is readily available on loading the package. The same data is also provided in TSV format. This is a limited subset of the complete data, and is included for package testing and exemplification purposes. The example discussed in Section \nameref{micompr:sec:examples:pphpc} uses a superset of this data, which is available for public download, but could not be included with the package due to its large size.

\subsection{Dependencies}

\pkg{micompr} has a number of optional dependencies, not required for package installation and for using most of its functionality. The \CRANpkg{biotools} \citep{silva2015biotools} and \CRANpkg{MVN} \citep{korkmaz2014mvn} packages are required by the \code{assumptions} functions, providing the statistical tests for assessing MANOVA and $t$-test assumptions. If these functions are invoked without the presence of the specified packages, they will inform the user of that fact, and terminate cleanly. The \CRANpkg{testthat} \citep{wickham2011testthat}, \CRANpkg{knitr} \citep{xie2015dynamic} and \CRANpkg{roxygen2} \citep{wickham2015roxygen2} packages are required for package development. The \CRANpkg{deseasonalize} package \citep{mcleod2013optimal} is required for building one of the vignettes.

\section{Examples}\label{micompr:sec:examples}

In this section we discuss four concrete application examples for the \pkg{micompr} package. The complete scripts used in these examples are available at \url{https://github.com/fakenmc/micompr-examples}.

\subsection{Simulation model with multiple outputs}\label{micompr:sec:examples:pphpc}

The replication of a simulation model in a new context highlights differences between the conceptual and implemented models, as well as inconsistencies in the conceptual model specification \citep{edmonds2003replication}, promoting model verification, model validation \citep{wilensky2007making}, and model credibility \citep{thiele2015replicating}. Some argue that a simulation model is untrustworthy until it has been successfully replicated \citep{edmonds2003replication,david2013validating}. Model parallelization is a an illustrative example of the importance of replication, as it is often required for simulating large models in practical time frames \citep{fachada2015parallelization}. By definition, model parallelization implies a number of changes, or even full reimplementation, of the original model, such that a robust comparison methodology, as provided by the \pkg{micompr} package, is required in order to make sure a parallelized model faithfully reproduces the behavior of the original serial model.

PPHPC is a reference model for studying and evaluating implementation strategies for spatial agent-based models, capturing important characteristics such as agent movement and local agent interactions \citep{fachada2015template}. The model describes a prototypical predator-prey system, and has six outputs, namely prey population, $P^s$, predator population, $P^w$, cell-bound food quantity, $P^c$, mean prey energy, $\mean{E}^s$, mean predator energy, $\mean{E}^w$, and mean cell-bound food levels, $\mean{C}$. Since outputs are collected once per iteration, each simulation run yields six time series, associated with the individual outputs. With several open source implementations publicly available, the model provides a good test case for multivariate comparison purposes.

Here we show the main comparison cases discussed in a previous article \citep{fachada2015model}, in which the model implementations are parameterized with size 400 and parameter set 1 \citep{fachada2015template}. A canonical PPHPC realization, implemented in NetLogo \citep{wilensky1999compat}, is compared with three configurations of a parallel Java implementation \citep{fachada2015parallelization}. The NetLogo implementation and the first Java configuration follow the PPHPC conceptual model and the specified parameters. The second Java configuration disables agent shuffling prior to agent actions, which is explicitly mandated in the conceptual model description. The third Java configuration performs a minimal change in one of the parameters specified by parameter set 1. As such, we define three comparison cases:

\begin{description}
\item[Case I] Compare the NetLogo implementation with the first Java configuration. These should yield distributionally equivalent results.
\item[Case II] Compare the NetLogo implementation with the second Java configuration. A small misalignment is to be expected.
\item[Case III] Compare the NetLogo implementation with the third Java configuration. There should be a mismatch in the outputs.
\end{description}

Independent samples of the six model outputs were obtained from $n=30$ replications for each implementation or configuration, in a total of $4n=120$ runs. Each replication $r=1,\ldots,4n$ was performed with a PRNG seed obtained by taking the MD5 checksum of $r$, guaranteeing independence between seeds, and consequently, between replications. The data generated by this computational experiment, as well as the scripts used to set up the experiment, are made available to other researchers at \url{https://zenodo.org/record/46848}.

The following script performs these comparisons. Note that the \code{concat = TRUE} option of the \code{micomp} function specifies that an additional concatenated output, $\widetilde{A}$, should be generated from the original outputs and analyzed in a similar fashion. The \code{dir\_data} variable specifies the location of the dataset.

\begin{example}
# Load package
library(micompr)

# Output names
outputs <- c("$P^s$", "$P^w$", "$P^c$", "$\\mean{E}^s$",
             "$\\overline{E}^w$", "$\\overline{C}$", "$\\widetilde{A}$")

# Outputs from the NetLogo implementation
dir_nl_ok <- paste0(dir_data, "nl_ok")
# Outputs from the Java implementation, first configuration
dir_jex_ok <- paste0(dir_data, "j_ex_ok")
# Outputs from the Java implementation, second configuration
dir_jex_noshuff <- paste0(dir_data, "j_ex_noshuff")
# Outputs from the Java implementation, third configuration
dir_jex_diff <- paste0(dir_data, "j_ex_diff")

# Files for model size 400, parameter set 1
filez <- glob2rx("stats400v1*.txt")

# Perform the three comparison cases
mic <- micomp(outputs,
              ve_npcs = 0.75,
              list(list(name = "I",
                        folders = c(dir_nl_ok, dir_jex_ok),
                        files = c(filez, filez),
                        lvls = c("NLOK", "JEXOK")),
                   list(name = "II",
                        folders = c(dir_nl_ok, dir_jex_noshuff),
                        files = c(filez, filez),
                        lvls = c("NLOK", "JEXNS")),
                   list(name = "III",
                        folders = c(dir_nl_ok, dir_jex_diff),
                        files = c(filez, filez),
                        lvls = c("NLOK","JEXDIF"))),
              concat = TRUE)
\end{example}            

\noindent The \code{mic} object can be inspected at the R prompt using the common S3 generic functions \code{print}, \code{summary} and \code{plot}. For publication purposes, the \code{toLatex} method for ``micomp'' objects produces \LaTeX\ tables with user\hyp{}specified information. For example, to generate a table similar to Table~4 of our previous work \citep{fachada2015model}, \code{toLatex} is invoked as follows:

\begin{example}
toLatex(mic, booktabs = TRUE,
        data_show = c("npcs-1", "mnvp-1", "parp-1", "scoreplot"),
        data_labels = c("$\\#$PCs", "MNV", "$t$-test", "PCS"),
        col_width = TRUE, pvalf_params = list(minval = 1e-8, na_str = "*"),
        label = "tab:pphpc",
        caption = paste("Comparison of a NetLogo implementation of",
                        "the PPHPC model against three configurations",
                        "of a parallel Java implementation."))
\end{example}

\noindent This call produces Table~\ref{tab:pphpc} with booktabs \citep{fear2005booktabsmanual} table style (\code{booktabs = TRUE}) and width set to document column width (\code{col\_width = TRUE}), since the table is somewhat large. The \code{label} and \code{caption} parameters set the label and caption of the \LaTeX\ table, respectively, while the \code{pvalf\_params} argument accepts a list of options for formatting $p$-values. The \code{data\_show} parameter specifies what data to show, which in this case is: 1) \code{npcs-1}, the number of PCs for the first specified variance (the \code{micomp} function accepts and performs output comparison with one or more specified variances); 2) \code{mnvp-1}, the MANOVA $p$-value for the first specified variance; 3) \code{parp-1}, $t$-test $p$-value for the first PC; and, 4) score plot for the first two PCs.

\begin{table}[ht]
\begin{center}
\resizebox{\columnwidth}{!}{%
\begin{tabular}{clrrrrrrr}
\toprule
\multirow{2}{*}{Comp.} & \multirow{2}{*}{Data} & \multicolumn{7}{c}{Outputs} \\
\cmidrule(l){3-9}
 &  & $P^s$ & $P^w$ & $P^c$ & $\mean{E}^s$ & $\mean{E}^w$ & $\mean{C}$ & $\widetilde{A}$\\
\midrule
\multirow{4}{*}{I}
 & $\#$PCs & 13 & 18 & 14 & 22 & 28 & 14 & 22\\
 & MNV & 0.325 & 0.640 & 0.462 & 0.052 & 0.796 & 0.463 & 0.523\\
 & $t$-test & 0.530 & 0.836 & 0.804 & 0.784 & 0.378 & 0.805 & 0.879\\
\bottomrule
\end{tabular}
} 
\caption{Comparison of a NetLogo implementation of the PPHPC model against three configurations of a parallel Java implementation.}
\label{tab:pphpc}
\end{center}
\end{table}

In terms of comparison, the method does not find significant differences in case I. However, it successfully differentiates the configurations compared in cases II and III. This is in line with what would be expected, and is discussed in further detail by \cite{fachada2015model}. While not shown here, the \code{assumptions(mic)} command reveals that most assumptions of the MANOVA and $t$-tests are verified.

\subsection{Monthly sunspots}\label{micompr:sec:examples:sunspots}

This example uses the monthly sunspot data \citep{sunspots}, included with R, which contains the monthly numbers of sunspots from 1749 to the present day. The solar cycle is an approximate 11-year period of changes in the number of sunspots and other associated phenomena. Thus, we divide the data into 11-year (132-month) periods, and consider each period to be an observation. In practice this is an oversimplification, since the cycles can be a bit longer or shorter than 11 years.

Given the data, we define two samples of 10 observations each, over a period of 110 years or 1320 months. The first sample includes solar cycles from 1749 to 1859, while the second encompasses cycles from 1902 to 2012. We can now ask the following question: were the solar cycles during the 1749--1859 interval significantly different from the more recent observations? The following code compares observations from the two time intervals, and attempts to provide an answer:

\begin{example}
# Load package
library(micompr)

# Months in the 1749-1859 interval (110 years)
# Months in the 1902-2012 interval (110 years)
m <- sunspot.month[c(1:1320, 1837:3156)]
m <- matrix(m, nrow = 20)

# Factor vector, two levels:
# a) ten 11-year cycles from 1749 to 1859
# b) ten 11-year cycles from 1902 to 2012
groups <- factor(c(rep("A", 10), rep("B", 10)))

# Compare the two groups, use 9 PCs for MANOVA
cmp <- cmpoutput("SunSpots", 9, m, groups)
\end{example}

\noindent The \code{cmp} object can now be analyzed:

\begin{example}
> cmp

Output name: SunSpots 
Number of PCs which explain 85.0
P-Value for MANOVA along 9 dimensions: 3.40755e-06
P-Value for t-test (1st PC): 2.713985e-06 
P-Value for Mann-Whitney U test (1st PC): 4.330035e-05 
Adjusted p-Value for t-test (1st PC): 6.513579e-06 
Adjusted p-Value for Mann-Whitney U test (1st PC): 0.0001039211 
\end{example}

\noindent The MANOVA $p$-value is significant, as well as the $t$-test and Mann-Whitney PC1 $p$-values, before and after weighted Bonferroni correction. As such, it is possible to conclude that solar cycles from 1749 to 1859 were significantly different from cycles between 1902 and 2012. However, is the data in accordance with the assumptions for the MANOVA and $t$-test? This can be checked with the \code{assumptions} function:

\begin{example}
> assumptions(cmp)
=== MANOVA assumptions ===
                  NPCs=9
Royston (A) 2.190796e-01
Royston (B) 3.627858e-01
Box's M     1.567180e-08

=== T-test assumptions ===
                       PC1
Shapiro-Wilk (A) 0.7739058
Shapiro-Wilk (B) 0.3791168
Bartlett         0.9353299
\end{example}

\noindent Only Box's M test, which checks for homogeneity of variance-covariance matrices, is significant. However, this test is prone to false positives, and this assumption is not critical when samples are of the same size \citep{tabachnick2013using}. Given this information, it seems plausible to consider the results provided by the parametric tests in our final decision, i.e., that there is in fact a significant difference between samples. A good way to visualize the overall results is to plot the ``cmpoutput'' object:

\begin{example}
> plot(cmp)
\end{example}

\begin{figure}[t!]
\centering
\includegraphics[width=\linewidth]{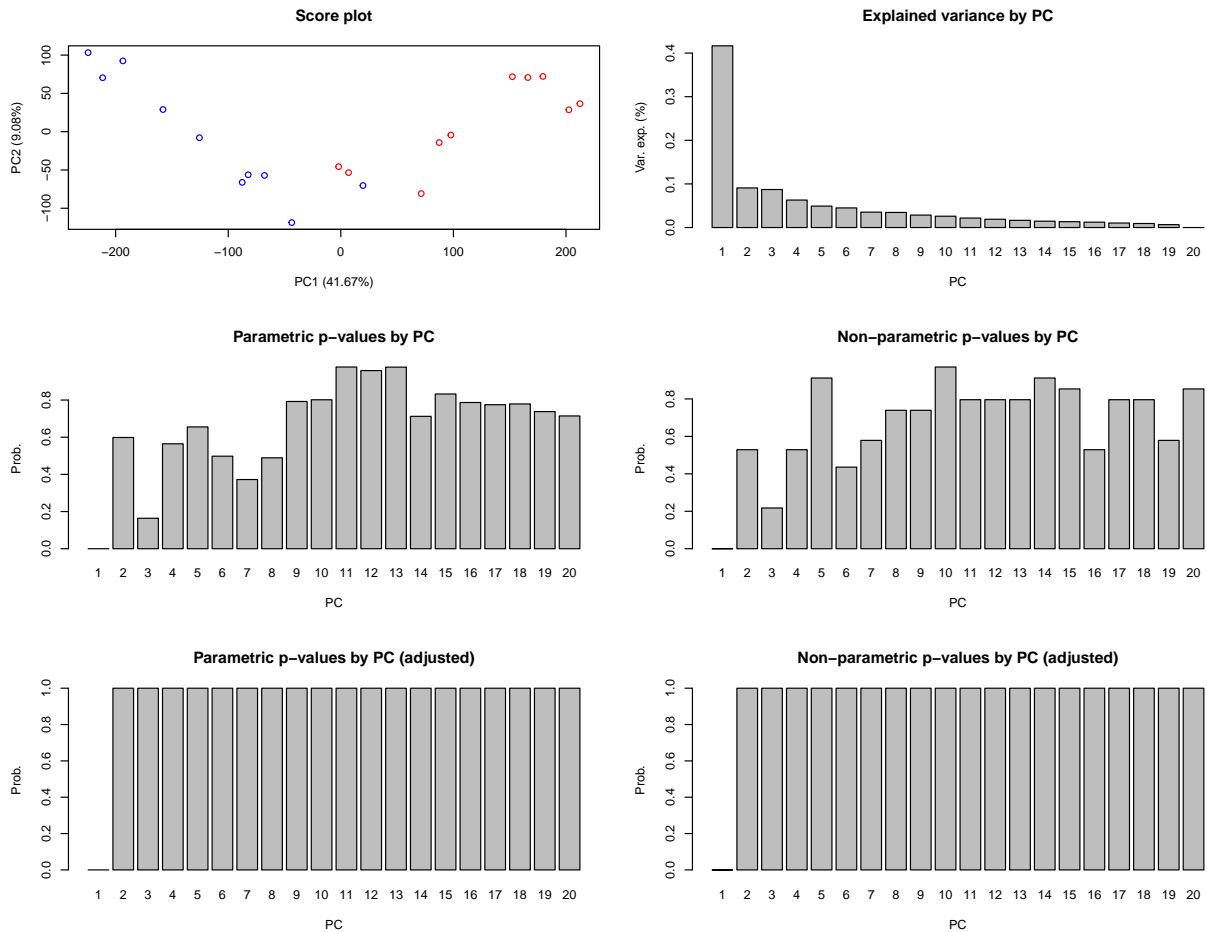}
\caption{Plots produced by sunspots example.}
\label{micompr:fig:sunspots}
\end{figure}

\noindent This command generates the plots shown in Figure \ref{micompr:fig:sunspots}. The score plot shows the samples to be distinctly separated, and the variance explained by PC decreases abruptly from the first PC to the second. Univariate $p$-values for PC1 are visibly significant, though not very much for the remaining PCs.

\subsection{Saugeen river flow}\label{micompr:sec:examples:saugeen}

This example uses the Saugeen River daily flow data \citep{hipel1994time}, included with the \pkg{deseasonalize} package. This data consists of a time series of the rivers' daily flow (m\textsuperscript{3}/s) from 1915 to 1979. Considering one year as an observation, there are a total of 65 observations. We can, for example, define two samples of 30 observations each, with the first and last 30 years of records, and ask the following question: is there any statistical difference between the flow dynamics during the 1915--1944 and 1950--1979 periods (perhaps due to climate change or some other factor)? The following code compares observations from the two periods: 

\begin{example}
# Load packages
library(micompr)
library(deseasonalize)

# Unique years
years <- unique(sapply(rownames(SaugeenDay), substr, 1, 4))

# Number of days in each year
ndays <- sapply(years, function(x) sum(substr(rownames(SaugeenDay), 1, 4) == x))

# Indexes of last day in each year
lastdays <- cumsum(ndays)

# Prepare data for PCA
saugdata <- t(mapply(
  function(nd, ld) {
    rflows <- rep(NA, 366)
    rflows[1:nd] <- SaugeenDay[(ld - nd + 1):ld]
    # Discard last day in leap years
    rflows[1:365]
  },
  ndays, lastdays))

# Consider first 30 years and last 30 years (discard 5 years in between)
saugdata <- saugdata[c(1:30, 36:65), ]

# Factor vector, two levels: first 30 years and last 30 years
groups <- factor(c(rep("A", 30), rep("B", 30)))

# Compare
cmp <- cmpoutput("SaugeenFlow", 0.9, saugdata, groups)
\end{example}

\noindent The \code{cmp} object can now be analyzed:

\begin{example}
> cmp
Output name: SaugeenFlow 
Number of PCs which explain 90.0
P-Value for MANOVA along 21 dimensions: 0.0740641
P-Value for t-test (1st PC): 0.3088125 
P-Value for Mann-Whitney U test (1st PC): 0.3980033 
Adjusted p-Value for t-test (1st PC): 1 
Adjusted p-Value for Mann-Whitney U test (1st PC): 1 
\end{example}

\noindent There are no significant $p$-values. As such, it is not possible to conclude that the river flow dynamic during the first 30 years of measurements is statistically different from the last 30 years. The MANOVA assumptions are not verified, as shown by be \code{assumptions} function:

\begin{example}
> assumptions(cmp)

=== MANOVA assumptions ===
                 NPCs=21
Royston (A) 1.309146e-05
Royston (B) 3.947504e-05
Box's M     1.859561e-10

=== T-test assumptions ===
                        PC1
Shapiro-Wilk (A) 0.02860029
Shapiro-Wilk (B) 0.83367671
Bartlett         0.79716243
\end{example}

\noindent The $t$-test assumptions mostly hold, in spite of the PC1 normality test for sample A (first 30 years) being significant at the $\alpha=0.05$ level. Nonetheless, the $U$ test $p$-value provides a similar conclusion. Plotting the \code{cmp} object provides another perspective:

\begin{example}
> plot(cmp)
\end{example}

\begin{figure}[t!]
\centering
\includegraphics[width=\linewidth]{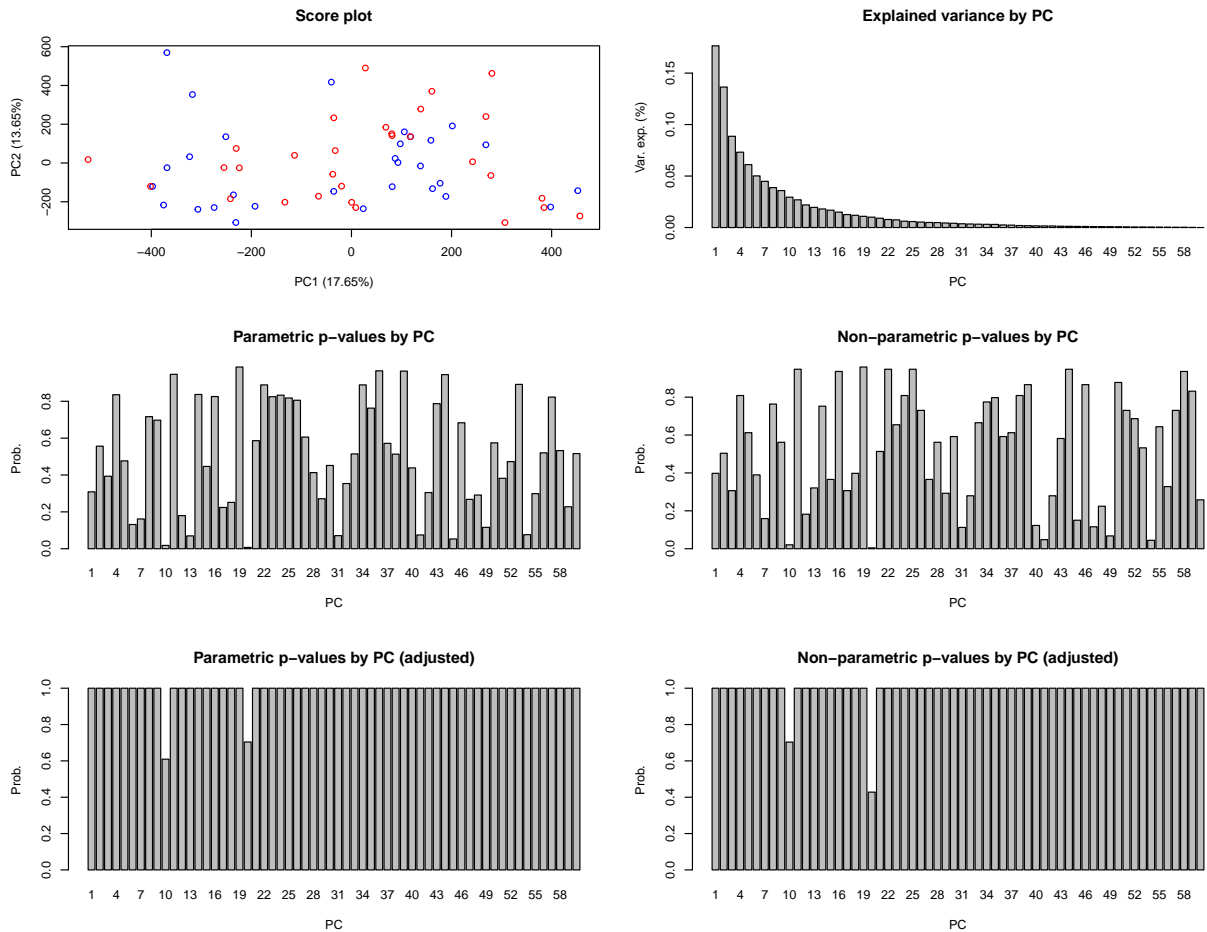}
\caption{Plots produced by the Saugeen river flow example.}
\label{micompr:fig:saugeen}
\end{figure}

\noindent The previous command will produce plots in Figure~\ref{micompr:fig:saugeen}. The PC1 vs. PC2 score plot does not show any clear sample separation and the decrease in explained variance from the first PC to the second is considerably less abrupt than what was observed for the sunspots example. Additionally, no $t$ and $U$ test $p$-values are significant for the 60 PCs after weighted Bonferroni correction, further reinforcing the conclusion that the yearly Saugeen river flow dynamic was similar during the compared periods of time.

\subsection{PH\textsuperscript{2} database of dermoscopic images}\label{micompr:sec:examples:ph2}

In this example we use the tools provided by the \pkg{micompr} package to study the PH\textsuperscript{2} database of dermoscopic images \citep{mendonca2013ph2}. This image database contains a total of 200 dermoscopic images of melanocytic lesions, including, from benign to more serious, 80 common nevi, 80 atypical nevi, and 40 melanomas. These are 8-bit RGB color images, with a resolution of purportedly $768 \times 560$ pixels. We have found, however, that resolutions vary between $761 \times 570$ and $769 \times 577$. As such, we resized all images to $760 \times 570$ prior to our analysis. The goal is to verify if images of the three types of lesions form statistically distinguishable samples, i.e., this is not a classification exercise such as performed by \cite{barata2014two}.

Each image is considered an observation of three outputs, \textbf{r}ed, \textbf{g}reen and \textbf{b}lue, corresponding to the respective color channels. The concatenation of all outputs, i.e., channels, provides a 4\textsuperscript{th} output. The three lesion samples are compared pairwise, as follows:

\begin{description}
\item[1v2] Common nevi and atypical nevi.
\item[1v3] Common nevi and melanomas.
\item[2v3] Atypical nevi and melanomas.
\end{description}

The following code reads the image dataset from disk and compares images grouped by lesion type. The \code{imgfolder} variable specifies the path containing the images (resized to $760 \times 570$), while the \code{grpsfile} variable specifies the path to the file containing the sample to which each image belongs.

\begin{example}

# Load packages
library(bmp)
library(micompr)

# Image definitions
imgs <- dir(imgfolder)
nimgs <- length(imgs)
npixels <- 760 * 570

# Specify image groups (Common nevi, atypical nevi, melanomas).
f <- read.table(grpsfile, row.names = 1)
grps <- f[order(row.names(f)), ]

# Read images from disk
# Use different color channels as outputs, and also use a concatenated output
rimgs <- matrix(nrow = nimgs, ncol = npixels)
gimgs <- matrix(nrow = nimgs, ncol = npixels)
bimgs <- matrix(nrow = nimgs, ncol = npixels)
rgbimgs <- matrix(nrow = nimgs, ncol = npixels * 3)

for (i in 1:nimgs) {
  cimg <- read.bmp(paste0(imgfolder, imgs[i]))
  rimgs[i, ] <- c(cimg[ , , 1])
  gimgs[i, ] <- c(cimg[ , , 2])
  bimgs[i, ] <- c(cimg[ , , 3])
  rgbimgs[i, ] <- c(cimg[ , , 1], cimg[ , , 2], cimg[ , , 3])
}

# Perform multivariate independent comparison of images
mic <-
  micomp(outputs = c("R", "G", "B", "RGB"),
         ve_npcs = 0.9,
         comps = list(
           list(name = "1v2",
                grpout = list(
                  data = list(R = rimgs[grps != 3, ],
                              G = gimgs[grps != 3, ],
                              B = bimgs[grps != 3, ],
                              RGB = rgbimgs[grps != 3, ]),
                  obs_lvls = factor(grps[grps != 3]))),
           list(name = "1v3",
                grpout = list(
                  data = list(R = rimgs[grps != 2, ],
                              G = gimgs[grps != 2, ],
                              B = bimgs[grps != 2, ],
                              RGB = rgbimgs[grps != 2, ]),
                  obs_lvls = factor(grps[grps != 2]))),
           list(name = "2v3",
                grpout = list(
                  data = list(R = rimgs[grps != 1, ],
                              G = gimgs[grps != 1, ],
                              B = bimgs[grps != 1, ],
                              RGB = rgbimgs[grps != 1, ]),
                  obs_lvls = factor(grps[grps != 1])))))
\end{example}

\noindent As in the \nameref{micompr:sec:examples:pphpc} example, the \code{mic} object can be inspected at the R prompt using the common S3 generic functions. Likewise, the \code{toLatex} function produces \LaTeX\ tables summarizing the object. The following code generates Table~\ref{tab:ph2}:

\begin{example}
toLatex(mic, booktabs = TRUE, data_show = c("parp-1", "nparp-1", "scoreplot"),
        data_labels = c("$t$-test", "$U$ test", "PCS"),
        pvalf_params = list(minval = 1e-8, na_str = "*"), label = "tab:ph2",
        caption = paste("Comparison of PH$^2$ dataset images",
                        "grouped by lesion type."))
\end{example}

\noindent Note that we did not request the MANOVA $p$-values in the \code{data\_show} parameter, as in this case the required assumptions do not appear to be verified. However, assumptions for the $t$-test on the first PC seem to hold. In any case, and to complement the information provided by the $t$-test, we specified the \code{"nparp-1"} option to the \code{data\_show} argument, such that the table shows the $p$-value of the Mann-Whitney $U$ test on the first PC.

\begin{table}[!t]
\begin{center}
\begin{tabular}{clrrrr}
\toprule
\multirow{2}{*}{Comp.} & \multirow{2}{*}{Data} & \multicolumn{4}{c}{Outputs} \\
\cmidrule(l){3-6}
 &  & R & G & B & RGB\\
\midrule
\multirow{3}{*}{1v2}
 & $t$-test & \uuline{2e-04} & \uuline{0.001} & \uline{0.033} & \uuline{0.002}\\
 & $U$ test & \uuline{8e-05} & \uuline{5e-04} & \uline{0.024} & \uuline{0.001}\\
\bottomrule
\end{tabular}
\caption{Comparison of PH$^2$ dataset images grouped by lesion type.}
\label{tab:ph2}
\end{center}
\end{table}

Results in Table~\ref{tab:ph2} show that images of different lesions have statistically significant differences, when compared either by individual color channels or with the three channels concatenated. The latter seems to provide better differentiation, with the common nevi and melanoma samples (1v3 comparison) appearing to be the most dissimilar.

\section{Summary}\label{micompr:sec:summary}

In this paper we presented the R package \pkg{micompr}, which implements a procedure for comparing multivariate samples associated with different factor levels or groups. The package architecture and its core components were discussed and four examples were examined.

\begin{rjonly}

\section*{Acknowledgments}

This work was supported by the Funda\c{c}\~{a}o para a Ci\^{e}ncia e a Tecnologia (FCT) projects UID/\allowbreak EEA/\allowbreak 50009/\allowbreak 2013 and UID/\allowbreak MAT/\allowbreak 04561/\allowbreak 2013, and partially funded with grant SFRH/\allowbreak BD/\allowbreak 48310/\allowbreak 2008, also from FCT.

\end{rjonly}

\begin{rjonly}

\end{rjonly}
\begin{arxivonly}

\end{arxivonly}

\begin{rjonly}

\address{Nuno Fachada\\
  	Institute for Systems and Robotics, LARSyS,\\
  	Instituto Superior T\'{e}cnico,\\
  	Universidade de Lisboa,\\
  	Lisboa,\\
  	Portugal}
\email{nfachada@laseeb.org}

\address{Jo\~{a}o Rodrigues\\
  	\'{E}cole Polytechnique F\'{e}d\'{e}rale de Lausanne,\\
  	Lausanne,\\
  	Switzerland}
\email{joao.rodrigues@epfl.ch}

\address{Vitor V. Lopes\\
  	UTEC - Universidad de Ingenier\'{i}a \& Tecnolog\'{i}a,\\
  	Lima,\\
  	Per\'{u}}
\email{vitorvieiralopes@gmail.com}

\address{Rui C. Martins\\
  	Life and Health Sciences Research Institute,\\
  	School of Health Sciences,\\
  	University of Minho,\\
  	Braga,\\
  	Portugal}
\email{rui.martins@ecsaude.uminho.pt}

\address{Agostinho C. Rosa\\
  	Institute for Systems and Robotics, LARSyS,\\
  	Instituto Superior T\'{e}cnico,\\
  	Universidade de Lisboa,\\
  	Lisboa,\\
  	Portugal}
\email{acrosa@laseeb.org}

\end{rjonly}


\begin{thebibliography}{43}
\providecommand{\natexlab}[1]{#1}
\providecommand{\url}[1]{\texttt{#1}}
\expandafter\ifx\csname urlstyle\endcsname\relax
  \providecommand{\doi}[1]{doi: #1}\else
  \providecommand{\doi}{doi: \begingroup \urlstyle{rm}\Url}\fi

\bibitem[Anderson(2001)]{anderson2001new}
M.~J. Anderson.
\newblock A new method for non-parametric multivariate analysis of variance.
\newblock \emph{Austral Ecology}, 26\penalty0 (1):\penalty0 32--46, Feb. 2001.
\newblock \doi{10.1111/j.1442-9993.2001.01070.pp.x}.

\bibitem[Barata et~al.(2014)Barata, Ruela, Francisco, Mendon{\c{c}}a, and
  Marques]{barata2014two}
C.~Barata, M.~Ruela, M.~Francisco, T.~Mendon{\c{c}}a, and J.~S. Marques.
\newblock Two systems for the detection of melanomas in dermoscopy images using
  texture and color features.
\newblock \emph{IEEE Systems Journal}, 8\penalty0 (3):\penalty0 965--979, Sept.
  2014.
\newblock \doi{10.1109/JSYST.2013.2271540}.

\bibitem[Baringhaus and Franz(2004)]{baringhaus2004new}
L.~Baringhaus and C.~Franz.
\newblock On a new multivariate two-sample test.
\newblock \emph{Journal of Multivariate Analysis}, 88\penalty0 (1):\penalty0
  190--206, jan 2004.
\newblock \doi{10.1016/s0047-259x(03)00079-4}.

\bibitem[Berg et~al.(2006)Berg, Hoefsloot, Westerhuis, Smilde, and
  Werf]{berg2006centering}
R.~A. Berg, H.~C. Hoefsloot, J.~A. Westerhuis, A.~K. Smilde, and M.~J. Werf.
\newblock Centering, scaling, and transformations: improving the biological
  information content of metabolomics data.
\newblock \emph{BMC Genomics}, 7\penalty0 (1):\penalty0 142, 2006.
\newblock \doi{10.1186/1471-2164-7-142}.

\bibitem[Clarke(1993)]{clarke1993nonparametric}
K.~R. Clarke.
\newblock Non-parametric multivariate analyses of changes in community
  structure.
\newblock \emph{Australian Journal of Ecology}, 18\penalty0 (1):\penalty0
  117--143, mar 1993.
\newblock \doi{10.1111/j.1442-9993.1993.tb00438.x}.
\newblock URL \url{http://dx.doi.org/10.1111/j.1442-9993.1993.tb00438.x}.

\bibitem[da~Silva(2016)]{silva2015biotools}
A.~R. da~Silva.
\newblock \emph{biotools: Tools for Biometry and Applied Statistics in
  Agricultural Science}, 2016.
\newblock URL \url{https://CRAN.R-project.org/package=biotools}.
\newblock R package version 3.0.

\bibitem[David(2013)]{david2013validating}
N.~David.
\newblock Validating simulations.
\newblock In \emph{Simulating Social Complexity}, Understanding Complex
  Systems, pages 135--171. Springer Berlin Heidelberg, 2013.
\newblock ISBN 978-3-540-93812-5.
\newblock \doi{10.1007/978-3-540-93813-2\_8}.

\bibitem[Duong(2016)]{duong2016ks}
T.~Duong.
\newblock \emph{ks: Kernel Smoothing}, 2016.
\newblock URL \url{https://CRAN.R-project.org/package=ks}.
\newblock R package version 1.10.4.

\bibitem[Duong et~al.(2012)Duong, Goud, and Schauer]{duong2012closed}
T.~Duong, B.~Goud, and K.~Schauer.
\newblock Closed-form density-based framework for automatic detection of
  cellular morphology changes.
\newblock \emph{Proceedings of the National Academy of Sciences}, 109\penalty0
  (22):\penalty0 8382--8387, may 2012.
\newblock \doi{10.1073/pnas.1117796109}.

\bibitem[Edmonds and Hales(2003)]{edmonds2003replication}
B.~Edmonds and D.~Hales.
\newblock Replication, replication and replication: Some hard lessons from
  model alignment.
\newblock \emph{Journal of Artificial Societies and Social Simulation},
  6\penalty0 (4):\penalty0 11, 2003.
\newblock URL \url{http://jasss.soc.surrey.ac.uk/6/4/11.html}.

\bibitem[Fachada et~al.(2015)Fachada, Lopes, Martins, and
  Rosa]{fachada2015template}
N.~Fachada, V.~V. Lopes, R.~C. Martins, and A.~C. Rosa.
\newblock Towards a standard model for research in agent-based modeling and
  simulation.
\newblock \emph{PeerJ Computer Science}, 1:\penalty0 e36, Nov. 2015.
\newblock \doi{10.7717/peerj-cs.36}.

\bibitem[Fachada et~al.(2016)Fachada, Lopes, Martins, and
  Rosa]{fachada2015parallelization}
N.~Fachada, V.~V. Lopes, R.~C. Martins, and A.~C. Rosa.
\newblock Parallelization strategies for spatial agent-based models.
\newblock \emph{International Journal of Parallel Programming}, pages 1--33,
  Jan. 2016.
\newblock \doi{10.1007/s10766-015-0399-9}.

\bibitem[Fachada et~al.(2017)Fachada, Lopes, Martins, and
  Rosa]{fachada2015model}
N.~Fachada, V.~V. Lopes, R.~C. Martins, and A.~C. Rosa.
\newblock Model-independent comparison of simulation output.
\newblock \emph{Simulation Modelling Practice and Theory}, 72:\penalty0
  131--149, Mar. 2017.
\newblock ISSN 1569-190X.
\newblock \doi{10.1016/j.simpat.2016.12.013}.
\newblock URL
  \url{http://www.sciencedirect.com/science/article/pii/S1569190X16302854}.

\bibitem[Fear(2005)]{fear2005booktabsmanual}
S.~Fear.
\newblock \emph{Publication quality tables in {LATEX}}, Apr. 2005.
\newblock URL \url{https://www.ctan.org/pkg/booktabs}.

\bibitem[Franz(2014)]{franz2014cramer}
C.~Franz.
\newblock \emph{cramer: Multivariate nonparametric {C}ramer-Test for the
  two-sample-problem}, 2014.
\newblock URL \url{https://CRAN.R-project.org/package=cramer}.
\newblock R package version 0.9-1.

\bibitem[Gibbons and Chakraborti(2010)]{gibbons2011nonparametric}
J.~D. Gibbons and S.~Chakraborti.
\newblock \emph{Nonparametric statistical inference}.
\newblock Statistics: Textbooks \& Monographs. Chapman and Hall/CRC, Boca
  Raton, FL, USA, fifth edition, July 2010.

\bibitem[Heller et~al.(2012)Heller, Small, and Rosenbaum]{heller2012crossmatch}
R.~Heller, D.~Small, and P.~Rosenbaum.
\newblock \emph{crossmatch: The Cross-match Test}, 2012.
\newblock URL \url{https://CRAN.R-project.org/package=crossmatch}.
\newblock R package version 1.3-1.

\bibitem[Hipel and McLeod(1994)]{hipel1994time}
K.~W. Hipel and A.~I. McLeod.
\newblock \emph{Time series modelling of water resources and environmental
  systems}.
\newblock Elsevier, 1994.

\bibitem[Jolliffe(2002)]{jolliffe2002principal}
I.~Jolliffe.
\newblock \emph{Principal component analysis}.
\newblock Springer Series in Statistics. Springer, second edition, 2002.
\newblock \doi{10.1007/b98835}.

\bibitem[Korkmaz et~al.(2014)Korkmaz, Goksuluk, and Zararsiz]{korkmaz2014mvn}
S.~Korkmaz, D.~Goksuluk, and G.~Zararsiz.
\newblock {MVN}: An {R} package for assessing multivariate normality.
\newblock \emph{The R Journal}, 6\penalty0 (2):\penalty0 151--162, 2014.
\newblock URL
  \url{http://journal.r-project.org/archive/2014-2/korkmaz-goksuluk-zararsiz.pdf}.

\bibitem[Kruskal and Wallis(1952)]{kruskal1952use}
W.~H. Kruskal and W.~A. Wallis.
\newblock Use of ranks in one-criterion variance analysis.
\newblock \emph{Journal of the American Statistical Association}, 47\penalty0
  (260):\penalty0 583--621, 1952.
\newblock \doi{10.1080/01621459.1952.10483441}.

\bibitem[Krzanowski(1988)]{krzanowski1998}
W.~J. Krzanowski.
\newblock \emph{Principles of Multivariate Analysis: A User's Perspective}.
\newblock Oxford University Press, New York, USA, 1988.

\bibitem[Massey~Jr.(1951)]{massey1951kolmogorov}
F.~J. Massey~Jr.
\newblock The {Kolmogorov-Smirnov} test for goodness of fit.
\newblock \emph{Journal of the American Statistical Association}, 46\penalty0
  (253):\penalty0 68--78, 1951.
\newblock \doi{10.1080/01621459.1951.10500769}.

\bibitem[McLeod and Gweon(2013)]{mcleod2013optimal}
A.~I. McLeod and H.~Gweon.
\newblock Optimal deseasonalization for monthly and daily geophysical time
  series.
\newblock \emph{Journal of Environmental Statistics}, 4\penalty0 (11), 2013.
\newblock URL \url{http://www.jenvstat.org/v04/i11}.

\bibitem[Mendon{\c{c}}a et~al.(2013)Mendon{\c{c}}a, Ferreira, Marques, Marcal,
  and Rozeira]{mendonca2013ph2}
T.~Mendon{\c{c}}a, P.~M. Ferreira, J.~Marques, A.~R.~S. Marcal, and J.~Rozeira.
\newblock {PH}\textsuperscript{2} - a dermoscopic image database for research
  and benchmarking.
\newblock In \emph{35th International Conference of the IEEE Engineering in
  Medicine and Biology Society (EMBC)}, pages 5437--5440. IEEE, July 2013.
\newblock \doi{10.1109/EMBC.2013.6610779}.

\bibitem[Mielke~Jr et~al.(1976)Mielke~Jr, Berry, and Johnson]{mielke1976multi}
P.~W. Mielke~Jr, K.~J. Berry, and E.~S. Johnson.
\newblock Multi-response permutation procedures for a priori classifications.
\newblock \emph{Communications in Statistics - Theory and Methods}, 5\penalty0
  (14):\penalty0 1409--1424, jan 1976.
\newblock \doi{10.1080/03610927608827451}.

\bibitem[Montgomery and Runger(2014)]{montgomery2010applied}
D.~C. Montgomery and G.~C. Runger.
\newblock \emph{Applied statistics and probability for engineers}.
\newblock John Wiley \& Sons, sixth edition, 2014.

\bibitem[Oksanen et~al.(2016)Oksanen, Blanchet, Friendly, Kindt, Legendre,
  McGlinn, Minchin, O'Hara, Simpson, Solymos, Stevens, Szoecs, and
  Wagner]{vegan2016}
J.~Oksanen, F.~G. Blanchet, M.~Friendly, R.~Kindt, P.~Legendre, D.~McGlinn,
  P.~R. Minchin, R.~B. O'Hara, G.~L. Simpson, P.~Solymos, M.~H.~H. Stevens,
  E.~Szoecs, and H.~Wagner.
\newblock \emph{vegan: Community Ecology Package}, 2016.
\newblock URL \url{https://CRAN.R-project.org/package=vegan}.
\newblock R package version 2.4-1.

\bibitem[{R Core Team}(2015)]{r2015stats}
{R Core Team}.
\newblock \emph{R: A Language and Environment for Statistical Computing}.
\newblock R Foundation for Statistical Computing, Vienna, Austria, 2015.
\newblock URL \url{https://www.R-project.org/}.

\bibitem[Rizzo and Szekely(2016)]{rizzo2016energy}
M.~L. Rizzo and G.~J. Szekely.
\newblock \emph{energy: E-Statistics: Multivariate Inference via the Energy of
  Data}, 2016.
\newblock URL \url{https://CRAN.R-project.org/package=energy}.
\newblock R package version 1.7-0.

\bibitem[Rosenbaum(2005)]{rosenbaum2005exact}
P.~R. Rosenbaum.
\newblock An exact distribution-free test comparing two multivariate
  distributions based on adjacency.
\newblock \emph{Journal of the Royal Statistical Society: Series B (Statistical
  Methodology)}, 67\penalty0 (4):\penalty0 515--530, sep 2005.
\newblock \doi{10.1111/j.1467-9868.2005.00513.x}.

\bibitem[Rosenthal and Rubin(1983)]{rosenthal1983ensemble}
R.~Rosenthal and D.~B. Rubin.
\newblock Ensemble-adjusted p values.
\newblock \emph{Psychological Bulletin}, 94\penalty0 (3):\penalty0 540--541,
  1983.
\newblock \doi{10.1037/0033-2909.94.3.540}.

\bibitem[Shaffer(1995)]{shaffer1995multiple}
J.~P. Shaffer.
\newblock Multiple hypothesis testing.
\newblock \emph{Annual Review of Psychology}, 46:\penalty0 561--584, Feb. 1995.
\newblock \doi{10.1146/annurev.ps.46.020195.003021}.

\bibitem[Sz{\'e}kely and Rizzo(2004)]{szekely2004testing}
G.~J. Sz{\'e}kely and M.~L. Rizzo.
\newblock Testing for equal distributions in high dimension.
\newblock \emph{InterStat}, 5:\penalty0 1--6, Nov. 2004.

\bibitem[Tabachnick and Fidell(2013)]{tabachnick2013using}
B.~G. Tabachnick and L.~S. Fidell.
\newblock \emph{Using multivariate statistics}.
\newblock Pearson, sixth edition, July 2013.

\bibitem[Talbert et~al.(2016)Talbert, Richards, Mielke, and
  Cade]{talbert2016blossom}
M.~Talbert, J.~Richards, P.~Mielke, and B.~Cade.
\newblock \emph{Blossom: Statistical Comparisons with Distance-Function Based
  Permutation Tests}, 2016.
\newblock URL \url{https://CRAN.R-project.org/package=Blossom}.
\newblock R package version 1.4.

\bibitem[Thiele and Grimm(2015)]{thiele2015replicating}
J.~C. Thiele and V.~Grimm.
\newblock Replicating and breaking models: good for you and good for ecology.
\newblock \emph{Oikos}, 124\penalty0 (6):\penalty0 691--696, 2015.
\newblock \doi{10.1111/oik.02170}.

\bibitem[WDC(2016)]{sunspots}
\emph{Sunspot Number}.
\newblock WDC-SILSO, Solar Influences Data Analysis Center (SIDC), Royal
  Observatory of Belgium, Brussels, Belgium, Apr. 2016.
\newblock URL \url{http://www.sidc.be/silso/datafiles}.

\bibitem[Wickham(2011)]{wickham2011testthat}
H.~Wickham.
\newblock testthat: Get started with testing.
\newblock \emph{The R Journal}, 3:\penalty0 5--10, 2011.
\newblock URL
  \url{http://journal.r-project.org/archive/2011-1/RJournal_2011-1_Wickham.pdf}.

\bibitem[Wickham et~al.(2015)Wickham, Danenberg, and
  Eugster]{wickham2015roxygen2}
H.~Wickham, P.~Danenberg, and M.~Eugster.
\newblock \emph{roxygen2: In-Source Documentation for {R}}, 2015.
\newblock URL \url{https://CRAN.R-project.org/package=roxygen2}.
\newblock R package version 5.0.1.

\bibitem[Wilensky(1999)]{wilensky1999compat}
U.~Wilensky.
\newblock \emph{{NetLogo}}.
\newblock Center for Connected Learning and Computer-Based Modeling,
  Northwestern University, Evanston, IL, USA, 1999.
\newblock URL \url{http://ccl.northwestern.edu/netlogo/}.

\bibitem[Wilensky and Rand(2007)]{wilensky2007making}
U.~Wilensky and W.~Rand.
\newblock Making models match: replicating an agent-based model.
\newblock \emph{Journal of Artificial Societies and Social Simulation},
  10\penalty0 (4):\penalty0 2, 2007.
\newblock URL \url{http://jasss.soc.surrey.ac.uk/10/4/2.html}.

\bibitem[Xie(2015)]{xie2015dynamic}
Y.~Xie.
\newblock \emph{Dynamic Documents with {R} and knitr}.
\newblock CRC Press, second edition, 2015.

\end{thebibliography}
\end{document}